\documentclass[11pt,a4paper]{article}

\RequirePackage[l2tabu, orthodox]{nag}

\usepackage{jheppub}
\usepackage{amsthm,amssymb,amsmath,epic,float}
\usepackage{rotating,epsfig,indentfirst,array,varioref}
\usepackage{appendix,tikz,mathtools}
\usepackage{setspace}
\usepackage{tikz}
\usetikzlibrary{decorations.pathreplacing}
\usetikzlibrary{calc}
\usepackage{enumitem}
\usepackage{hyphenat}
\usetikzlibrary{positioning}
\usepackage{graphicx}  
\usepackage{adjustbox}

\usepackage{arydshln}

\def\bbt{\bibitem}
\def\be{\begin{equation}}
\def\en{\end{equation}}
\def\ber{\begin{eqnarray}}
\def\enr{\end{eqnarray}}

\def\d{\partial}

\def\al{\alpha}

\def\im{\imath}

\def\lm{\lambda}
\def\Lm{\Lambda}

\def\om{\omega}

\def\eps{\epsilon}

\def\fr12{\frac{1}{2}}

\usepackage{jheppub}
\makeatletter
\def\@fpheader{\vspace{-.1cm}}
\makeatother

\title{\boldmath  
Free field construction of Heterotic string compactified on Calabi-Yau manifolds of Berglund-Hubsch type in  the Batyrev-Borisov combinatorial approach\\}

\author{Alexander Belavin,}

\affiliation{Landau Institute for Theoretical Physics, 142432 Chernogolovka, Russia}

\emailAdd{belavin@itp.ac.ru}

\abstract{Heterotic string models in $4$-dimensions are the hybrid theories of a left-moving $N=1$ fermionic string  whose additional $6$-dimensions are compactified on a  $N=2$ SCFT theory with the central charge $9$, and a right-moving bosonic string, whose additional  dimensions are also compactified on $N=2$ SCFT theory with the central charge $9$,  and the remaining $13$ dimensions compactified on the torus of $E(8)\times SO(10)$ Lie algebra.

The important class of exactly solvable Heterotic string models considered earlier by D. Gepner corresponds to the products of $N=2$ minimal models with the total central charge $c=9$. These models are known to describe Heterotic string models compactified on Calabi-Yau manifolds, which belong a special subclass of general CY manifolds of Berglund-Hubsch type. We generalize this construction to all cases of compactifications on Calabi-Yau manifolds of general Berglund-Hubsch type, using Batyrev-Borisov combinatorial approach. In particular, starting from the mirror pair of Batyrev lattices corresponding to a given CY manifold, we construct vertex operators of the complete physical theory as cohomology of Borisov differentials that correspond to points of reflexive Batyrev polyhedra. In particular, we show how the number of $27$, $\overline{27}$  and Singlet representations of $E(6)$ is determined by the data of reflexive Batyrev polytope that determines this CY-manifold.}


\begin{document}
\maketitle

\section{Introduction}\label{sec:1}


Heterotic string  approach to unifying Quantum gravity theory and Grand Unified Gauge Theory in 10 dimensions was proposed in \cite{GHMR,CHSW}. In  \cite{Gep}  (see also \cite {BLT}) a conjecture was put forward about the equivalence of the compactification of $6$ dimensions of $10$-dimensional spacetime on Calabi-Yau manifolds and the compactification on $N=2$ SCFT with total central charge $c=9$.

The construction of Heterotic string Gepner models starts as a product of two CFTs with the critical central charges. Namely, they are the product of a left-handed fermionic string obtained from an $N=1$ CFT whose extra 6 dimensions are compactified into a product of minimal $N=2$ SCFT models with a total central charge of $9$, and a right-handed bosonic string whose extra $9$ dimensions are also compactified onto a product of minimal models.
The remaining $13$ right dimensions are compactified on the torus of the algebra $E(8)\times SO(10)$.

As a result of the construction, these models obtain the space-time $N=1$ supersymmetry, arising from the GSO reduction on the left side, and the gauge symmetry $E(8)\times E(6)$, arising from a similar reduction on the right side.

In this work, we  propose a method for  construction the  models of Heterotic string compactified on the product of a torus of  $E(8)\times SO(10)$  Lie algebra  and on Calabi-Yau manifolds of general Berglund-Hubsch type \cite{Berglund:1991pp,Kreuzer:2000xy,Krawitz}.  The method uses the Batyrev-Borisov combinatorial approach \cite{BAT}, to implement a vertex algebra realized by free bosonic and fermionic fields for the states of the Calabi-Yau sector. We stress that this method is similar to the well known free-field representation \cite{Felder:1988zp} of $N=0$ minimal models \cite{BPZ}. In particular, it involves screening operators and the corresponding BRST like complexes that reduce the number of physical states. The screening operators (Borisov differentials $D_{\vec{m}}$ and $D_{\vec{n}}$) correpond to the points of Batyrev reflexive polyhedra.

The construction uses  the requirement for the simultaneous fulfillment of mutual locality of the left-moving  vertices  with the $N=1$ SUSY  generators and of right-moving vertices with generators of $E(8)\times E(6)$-gauge symmetry. The $78$ $E(6)$ currents apprear as a result of extending of $SO(10)$ currents ($45$ currents of adjoint representation) by the vertex operators corresponding to two $SO(10)$ spinor representations dressed by appropriate $U(1)$ factors originating from CY sector ($32=16+16$ currents) and an additional purely CY $U(1)$ current ($1$ additional current). We empasize that the phenomenon of $E(6)$ enhancement and interplay with $N=1$ supersymmetry is well known in the literature (see \cite{Lerche:1988np} for review).

The further requirement of mutual locality of the products of the left and right vertices of physical states with each other and together with other requirements of the Conformal bootstrap \cite{BPZ} (see also \cite{BEL1,BEL2}), leads to a self-consistent result precisely for the type of CY chosen above. Based on these requirements, we explicitly construct Vertex operators of the physical states of the theory in the following order.

We first present Borisov's construction for $N=2$ SCFT models corresponding to Calabi-Yau manifolds of Berglund-Hubsch type. Then, in the  left moving sector of the model  we find the generators  of  $N=1$ spacetime supersymmetry  and choose the set of left-movers that are mutually local with them. After this, in the  right sector we find the additional generators of $E(6)$ symmetry and choose the set of right movers  that are mutually local with them. 

At the last step, we find the products of the left- and right-moving vertices of the states  obtained in this way, which are mutually local to each other and permit to achieve modular invariance in constructing heterotic string theory compactified on CY-manifolds of BH type.

\section{The original Conformal field theory}\label{sec:2}

The construction of the Heterotic string starts with a theory that is the product of two Conformal Field theories: $N=1$ SCFT on the left, holomorphic side and $N=0$ CFT on the right, antiholomorphic side.

In turn, in the left sector we have the product of the 4-dimensional $N=1$ CFT for the subsector of space-time with central charge 6, 
and the $N=1$ CFT for the compact Calabi-Yau subsector with central charge $9$, so that the total central charge in the left sector is $c =15$.
  
The $N=1$ SCFT of the left space-time subsector is a theory of $4$ free bosons $x^{\mu}(z)$ and $4$ Majorana fermions $\psi^{\mu}(z)$.
	
The $N=1$  SCFT of the left  Calabi-Yau subsector with central charge 9 is realized in accordance with the work of L. Borisov using the theory of free bosons $X^{\pm}_i(z)$ and Majorana fermions $\Psi^{\pm}_i(z)$, where $i=1,...,5$. The algebra of vertex operators in this subsector is determined by the data of a Calabi-Yau manifold of Berglund-Hubsch type. We formulate how below. Left-moving energy-momentum tensor is 
\begin{equation}
	T^L(z)=-\frac{1}{2}\eta_{\mu\nu}\d x^{\mu}\d x^{\nu} 
	     - \frac{1}{2}\eta_{\mu\nu}\psi^{\mu}(z) \d\psi^{\nu}(z)+T^L_{CY}(z).
\end{equation}
	
In the right sector we have the product of  the  CFT  for the space-time subsector with central charge $4$, realized by $4$ bosonic fields $\bar{x}^{\mu}(\bar{z})$, the CFT for a subsector with the central charge $8$, realized by $8$ free boson fields $Y_{I}(\bar{z}), I=1,...,8$ compactified on the torus of the $E(8)$ algebra, the  CFT for the subsector with the central charge $5$, realized by $5$ free bosonic fields $\Phi_{\alpha}(\bar{z}), \alpha=1,...,5$ compactified on the torus of the $SO(10)$ algebra, and the CFT of the compact Calabi-Yau subsector with central charge $9$, so that the total central charge in the right sector is $c =26$.

The CFT of the right Calabi-Yau subsector with the central charge $9$ is realized similarly to the left one by free bosons $\bar{X}^{\pm}_i(\bar{z})$ and 4 Majorana fermions $	\bar{\Psi}^{\pm}_i(\bar{z})$,  where $i=1,...,5$.

Right-moving energy-momentum tensor is 
\begin{equation}
	T^R(\bar{z})=-\fr12\eta_{\mu\nu}\bar{\d}\bar{x}^{\mu}\bar{\d}\bar{x}^{\nu}+
	\fr12(\bar{\d}Y_{I})^{2}+\fr12(\bar{\d}\Phi_{\alpha})^{2}+ T^R_{CY}(\bar{z}).
\end{equation}
	
\section{Calabi-Yau manifolds and N=2 SCFT}\label{sec:3}	
In the works \cite{BOR1,BOR2} L. Borisov formulated the construction of N=2 SCFT models corresponding to Calabi-Yau models.  The $N=2$ Virasoro superalgebra generators in this construction are expressed in terms of the free bosons $X^{\pm}_i(z)$,   the Majorana fermions $\Psi^{\pm}_i(z)$, where $i=1,...,5$.  We will also use the free boson fields $H_i,i=1,...,5$, in terms of which the Majorana fermions are expressed as
$\Psi^{\pm}_i(z)=\exp{[\pm i H_i(z)]}$.

The operator product expansion (OPE) of these fields looks as follows
\begin{equation}\label{2.XpsiOPE}
\begin{aligned}
	&X^{+}_i(u)X^{-}_j(z)=\delta_{ij}\log(u-z)+\dots, \\ 
	&\Psi^{+}_i(u)\Psi^{-}_j(z)=\delta_{ij} (u-z)^{-1}+\dots,\\
	&H_i(u) H_j(z)=-\delta_{ij}\log(u-z)+\dots
\end{aligned}
\end{equation}

The currents $T_{CY}(z)$, $G^{\pm}_{CY}(z)$ and  the $U(1)$ current $J_{CY}(z)$ that form $N=2$ Virasoro algebra of Calabi-Yau subsectors look as follows
\begin{equation}
\begin{aligned}
	&T_{CY}(z)= \sum_{i=1}^{5}[\d X^{+}_i \d X^{-}_i+
	\frac{1}{2}(\d\Psi^{+}_i\Psi^{-}_i+\d\Psi^{-}_i\Psi^{+}_i)-
	\frac{1}{2} (a^+_i \d^2 X^{-}_i + a^-_i \d^2 X^{+}_i)],\\
	&G^{+}_{CY}(z)=\sqrt{2} \sum_{i=1}^{5}
	\left(\Psi^{+}_i\d X^{-}_i- \d\Psi^{+}_i a^-_i \right),\\
	&G^{-}_{CY}(z)=\sqrt{2} \sum_{i=1}^{5}
	\left(\Psi^{-}_i\d X^{+}_i-\d\Psi^{-}_i a^+_i \right),\\
	&	J_{CY}(z)=\sum_{i=1}^{5}
	 \left(\Psi^{+}_i\Psi^{-}_i -a^+_i \d X^{-}_i +a^-_i \d X^{+}_i\right).
\end{aligned}
\end{equation}

It is also be useful to represent the currents $T_{CY}(z)$ and $J_{CY}(z)$ 
in the following equivalent bosonic form
\be
\begin{aligned}
	&T_{CY}(z)= \sum_{i=1}^{5}\left[\d X^{+}_i \d X^{-}_i-
	\frac{1}{2} (\d H_i)^2 -
	\frac{1}{2}\left(a^+_i \d^2 X^{-}_i + a^-_i \d^2 X^{+}_i\right)\right],\\
	&J_{CY}(z)= \d \sum_{i=1}^{5}\left[iH_i-a^+_i X^{-}_i + a^-_i X^{+}_i \right]=\d H_{CY}(z).
\end{aligned}
\en

The vectors $\vec{a}^+$ and $\vec{a}^-$ depend on the Calabi-Yau manifold selected for  compactification on it.   Below we  will show how to do this , and also that the scalar product of these vectors  $(\vec{a}^+,\vec{a}^-)$ is equal $1$. The central charge $c$ of the $N=2$ Virasoro algebra under consideration is expressed through these vectors as follows
\begin{equation}
\frac{c}{3}=5-2\sum_{i=1}^{5}a^+_i a^-_i. 
\end{equation}
These statements  will be explained below. Thus, the central charge of the left and right Calabi-Yau subsectors is $9$, which is necessary for  Heterotic string theory to be self-consistent.
\section {Berglund-Hubsch  type CY manifolds and N=2 SCFT models}
\label{sec:4}
The purpose of this section is to briefly remind  the  Borisov  construction the lattice vertex algebra  of Berglund-Hubsch mirror symmetry  with the $N=2$ superconformal structure. 

Calabi-Yau manifolds of BH type \cite{Berglund:1991pp,Krawitz} are defined as a hypersurface in the weighted projective space $P_{\vec{k}}$ or in its orbifold defined by the equation
\begin{equation}
W(y_1,...,y_5)=\sum_{i=1}^{5}\prod_{j=1}^5 y_j^{A_{ij}}=0.
\end{equation} 
Here $\vec{k}=k_1,...,k_5$, where $ k_i,i=1,...,5$ are the weights  
of $P_{\vec{k}}$.
$W(y_ 1,...,y_5)$ is a non-degenerate polynomial  with invertible  integer matrix $A_{ij}$.
Non-degeneracy of the potential means that the hypersurface $W =0$ in $P_{\vec{k}}$  is smooth away from the origin.  It is a very restrictive  condition, and complete classification of non-degenerate potentials is  given in \cite{Kreuzer:2000xy}.
 
It is assumed that the variables $y_i$ have positive rational degrees 
$q_i=\frac{k_i}{d}$,  and $d=\sum_{i=1}^5 k_i$.

This means that $\sum A_{ij}q_j = 1$ for all $i$, and the polynomial $W$  as well as a monomial  $\prod_{i=1}^5 y_i$ are homogeneous with respect to  the replacement $y_j\rightarrow\exp(i 2\pi q_j)y_j$.
  
The symmetry group $G$ described above always exists, since it simply follows from the self-consistency between the data of the projective space $P_{\vec{k}}$ and the polynomial $W$.  Let's call it the "minimum admissible group." The maximum allowed group can be larger \cite{Krawitz}. How much exactly depends on the projective space of the data. But below, for simplicity, we will use the case of the minimal admissible group when constructing the initial CY manifold.

What about the mirror CY-manifold, which is defined by the mirror polynomial  $W^T$ with the transposed matrix $A^T_{ij}$ in the mirror projective space  $P_{\vec{k}}^*$, and by the dual admissible group $G^T$, this group will be determined automatically from the duality requirements.

From the described Berglund-Hubsch data, the potential $W$ and the initial admissible group $G$, one can obtain the Batyrev-Borisov combinatorial data, which will be used in constructing the Calabi-Yau sectors in the Heterotic string.
   
Let $M_0$ and $N_0$ be two integer $5$-dimensional lattices with bases  $\vec{u_i}$ and $\vec{v_j}$, whose pairing by definition looks like this 
\begin{equation}
\vec{u_i}\cdot\vec{v_j} =A_{ij},
\end{equation}
where $A_{ij}$ are the exponents of the potential $W$.

The solution to these equations can be conveniently chosen as follows
\begin{equation}
(\vec{u_i})_j =A_{ij}, \ \ (\vec{v_i})_j =\delta_{ij}.
\end{equation}
We also define two vectors $\vec{a}^+$ and $\vec{a}^-$,
which are related to the vectors $\vec{u_i}$ and $\vec{v_j}$
as follows
\begin{equation}
\vec{a}^+ =\frac{1}{d^*}\sum_{i} k^*_i \vec{u}_i,  \ \
\ \vec{a}^- = \frac{1}{d}\sum_{j} k_j\vec{v_j}.
\end{equation}
It is easy to  verify that these vectors satisfy to the following equations 
\begin{equation}
\vec{u}_i\cdot \vec{a}^- = 1,  \  \  \vec{a}^+ \cdot \vec{v}_j=1.
\end{equation}

Now we define an extension of the lattices $M_0$ and $N_0$ to a pair of dual Batyrev  lattices $M$ and $N$ for the case when the admissible group $G$ for the original polynomial $W$ is chosen to be minimal.

First, to  perform this we  extend the lattice $N_0$ to the lattice
$N=N_0 + \vec{a}^-$ as in \cite{BOR1,BOR2}. In the next step we find the basis of the lattice $N$. We denote its elements as $\vec{e^*}_{\beta=1,...,5}$.
Then we find the basis of the dual lattice $M$ as the five vectors
$\vec{e}_{\alpha}$ whose pairing with  $\vec{e}^*_{\beta}$, ($\alpha, \beta=1,...,5$) is 
\begin{equation}
\vec{e}_{\alpha} \cdot \vec{e^*}_{\beta}=\delta_{\alpha,\beta}.
\end{equation}
	
The important role  in the construction of Heterotic strings compactified on Calabi-Yau manifolds of the Berglund-Hubsch type will be played by the set of elements $\vec {m}\in M$ and $\vec {n}\in N$, which are combinations of bases $M$ and $N$ with non-negative coefficients and belonging to the reflexive Batyrev polytopes $\Delta^+$ and $\Delta^-$.	
The latter means  means that $\vec {m}\in \Delta^+$ and $\vec{n}\in \Delta^-$, if
\begin{equation}
\vec{m} \cdot \vec{a}^- = 1, \  \   \vec{a}^+ \cdot \vec{n}=1.
\end{equation}
This equality, as shown above, ensures that the magnitude of the central charge in the subsector CY is equal to $9$.

In constructing the vertex algebra of the left and right CY-subsectors of  Heterotic string, following the work of Borisov \cite{BOR1,BOR2}, we use vertices corresponding to the points of the dual lattices $M$ and $N$.
In fact, this is only true for vertices that correspond to bosonic states (in the left sector, they correspond to NS states).
To construct fermion states  we need to use in both sectors, left and right, vertices corresponding to the following extension of the $M$ and $N$ lattices, namely, $M \pm \frac{1}{2}\vec{a^+}$ and $N \pm \frac{1}{2}\vec{a^-}$.

The reason for this extension is that we want to build not just a SCFT model,  but Heterotic string theory that includes SCFT of CY as subsector. Therefore, since we have a diagonal $N=1$ SCFT in the left sector (and also in the space-time and CY subsectors), then in both left subsectors the vertex operators must simultaneously belong to either the Ramond (R) or the Neveu-Schwatz (NS) type. 

For the theory to be consistent, we need both options.  Moreover, in the absence of vertex algebras of both types (NS-type and R-type), we cannot obtain space-time SUSY. A similar situation occurs in the right sector. This will be explained below when we construct vertex algebras first in the left and then in the right sectors.

Namely, following  \cite{BOR1,BOR2}, we define the Vertex algebra of Heterotic string as cohomology with respect to the  sum of differentials $D_{\vec{u}_i}$ and $D_{\vec{v}_j}$
\begin{equation}\label{Borisov-differentials}
D=\sum_{i=1}^5D_{\vec{u}_i}+\sum_{j=1}^5D_{\vec{v}_j},
\end{equation}
where
\be
\begin{aligned}
&D_{\vec{u}_i}=\oint dz \  \vec{u}_i \cdot \vec{\Psi}^-(z) \  
\exp (\vec{u}_i \cdot \vec{X^-}(z)),\\
&D_{\vec{v}_j}=\oint dz \ \vec{v}_j \cdot \vec {\Psi}^+(z) \ 
\exp (\vec{v}_j \cdot \vec{X^+}(z)).
\end{aligned}
\en
It was stated in \cite{BOR1,BOR2} that the elements $T$,  $G^{\pm}$ and $J$ belong to the kernel of the operator $D$.

Calabi-Yau multiplier of any total vertex operator of  Heterotic string contains an exponential factor of the following form
\begin{equation}
\exp\left(i\sum_{i=1}^5 S_iH_i+\vec{p} \cdot \vec{X}^{-}+\vec{q} \cdot \vec{X}^{+}\right),
\end{equation}
whose dimensions is 
\begin{equation}
\Delta(\vec{S},\vec{p},\vec{q})=
\frac{1}{2}\vec{S}^2 + \vec{p} \cdot \vec{q}+
\frac{1}{2}\left(\vec{p} \cdot \vec{a}^- +  \vec{q} \cdot \vec{a}^+\right).
\end{equation}

These CY factors of the vertex operators have the following properties.

Firstly,  their  parameters $\vec{S}, \ \vec{p}, \ \vec{q}$ must be elements  of three $5$ dimensional lattices to satisfy the OPE axioms. 

Secondly, their dimensions must be equal, as we see,  to $0$ or $1/2$ modulo an integer in the NS-case, or equal to $3/8$ modulo an integer in the R-case.

Thirdly, they must  be consistent or, more precisely, be mutually local with the differentials $D_{\vec{m}}$ and $D_{\vec{m}}$. For this, the parameters $\vec{S}, \ \vec{p}, \  \vec{q}$ must satisfy the requirement that the pairing $(\vec{p},\vec{a}^-)$ and $(\vec{a}^+,\vec{n})$ be integers  or half -integers for all $\vec{p}$ and $\vec{q}$.

It follows that if the vertex operator belongs to the NS-sector, then all $S_i$ are integers, $\vec {p}\in M$ and $\vec{q}\in N$, and if the vertex operator belongs to the R-sector, then  all $S_i$ are half-integers, $\vec {p} \in M \pm \frac{1}{2}\vec{a}^+$ 
and $\vec{q} \in N \pm \frac{1}{2}\vec{a}^{-}$.

For shortness, we will call the set of states belonging to the intersection of the set of BRST cohomology defined by the differential $Q_{BRST}$ and the set of  Borisov cohomology \cite{BOR1,BOR2} defined by the differential $ D_{\vec{1},\vec{1}}$, the space of quasi-physical states.

However, below we find two additional reductions of this set, which lead to an extension in the left sector of Poincaré symmetry to $N1$ SUSY spacetime, and in the right sector to an extension of $E(8)\times SO(10)$ to $E(8)\times E(6)$.

It is this set that will be the space of physical states of the Heterotic string theory.

\section{Left-moving sector, N=1  SCFT}\label{sec:5}
In the left sector we have the product of  4-dimensional $N=1$ CFT for the left  space-time subsector consisting of $4$ bosons and $4$ Majorana fermions  with the central charge of 6 and  $N=1$ CFT for the compact subsector with the central charge of 9, so that the total central charge in the left sector is $c =15$.

The $N=1$ CFT of the left space-time factor is a theory of 4 free boson fields  
$x^{\mu}(z)$ and 4 Majorana fermion fields $\psi^{\mu}(z)$
\be
\begin{aligned}
	&x^{\mu}(z)x^{\nu}(0)=-\eta^{\mu\nu}\log{z}+\dots,\\
	&\psi^{\mu}(z)\psi^{\nu}(0)=\eta^{\mu\nu}z^{-1}+\dots,
	\label{2.XpsiOPE}
\end{aligned}
\en

As for the $N=1$ CFT of the left Calabi-Yau subsector, this is the above-described theory of free bosons $X^{\pm}_i(z)$, fermions $\Psi^{\pm}_i(z)$ and free boson fields $H_i,i=1,...,5$ bosonizing the fermions.  Here the $N=1$ symmetry corresponds to the subalgebra of the $N=2$, which was  defined above when representing the CY subsectors.

The total left-moving $N=1$ Virasoro algebra is the diagonal subalgebra in the direct sum  of the  Calabi-Yau  compact subsector, introduced above and   of the $N=1$ Virasoro algebras of space-time degrees of freedom
\begin{equation}
	\begin{aligned}
		&T^L(z)=T^L_{ST}(z)+T^L_{CY}(z), 
		\\
		& G^L(z)=G^L_{ST}(z)+G^L_{CY}(z), 
		\\
		&T^L_{ST}=-\frac{1}{2}\d x^{\mu}(z) \d x_{\mu}(z)-
		\frac{1}{2}\psi^{\mu}(z)\d\psi_{\mu}(z), 
		\\
		&G_{ST}(z)=\d x^{\mu}\psi_{\mu}(z),
		\\
		\
		&G^L_{CY}(z)=G^{+}_{CY}+G^{-}_{CY}(z).
		\label{LVir}
	\end{aligned}		
\end{equation}

The $N=1$ Virasoro superalgebra action is correctly defined on the product of only $NS$-representations or on the product of only $R$-representations.

We use BRST approach to define the physical states. The BRST charge is given by the integral 
\begin{equation}
	Q_{\textrm{BRST}}=\oint dz 
	[cT_{mat}+\gamma G_{mat}+\fr12(cT_{gh}+\gamma G_{gh})],
	\label{BRST}
\end{equation}
where we introduced the  ghost fields and $N=1$ Virasoro superalgebra of the ghosts
\begin{equation}
	\beta(z)\gamma(0)=-z^{-1}+..., \ \ b(z)c(0)=z^{-1}+....
	\label{ghope}
\end{equation}
\begin{equation}
	\begin{aligned}
		&T_{\textrm{gh}}=-\d bc-2b\d c-\frac{1}{2}\d\beta\gamma-\frac{3}{2}\beta\d\gamma,
		\\
		&G_{\textrm{gh}}=\d\beta c+\frac{3}{2}\beta\d c-2 b\gamma.
		\label{1.Virgh}
	\end{aligned}
\end{equation}
The ghost  space of states is characterized by the vacuum $V_{q}(z)$, which can be realized as a free scalar field exponent
\begin{equation}
	\begin{aligned}
		&V_{q}(z)=\exp{(q\phi(z))},\
		\phi(z)\phi(0)=-\log(z)+\dots
	\end{aligned}
\end{equation}

The left-moving vertex can be written as
\begin{multline}
V^L_{\vec{\mu}}=
		P(\d^k x^{\mu},\d^l \tilde{H}_{a}, \d^r H_{i},\d^s X^+_{i},\d^t X^-_{i})\times\\\times
			\exp\left(q\phi+ \im\lm^{a}\tilde{H}_{a}+
			i\sum_{i=1}^5 S_iH_i+\vec{p}\vec{X}^-+\vec{q}\vec{X}^{+}+ \im p_{\mu}x^{\mu}(z)\right). 
\end{multline}

Here $P$ is a polynomial of the  derivatives  of the corresponding boson fields including the fields $\tilde{H}_a, a=1,2$. These fields bosonize the Fermi fields of spacetime  subsector as follows  
\begin{equation}
\begin{aligned}
&\tilde{H}_a(z)\tilde{H}_b(0)=-\delta_{ab}\log{(z)}+..., \ a,b=1,2.
	\\
	&\frac{1}{\sqrt{2}}(\pm\psi^{0}+\psi^{1})=\exp{[\pm\im \tilde{H}_{1}]},\
	\frac{1}{\sqrt{2}}(\psi^{2}\pm\im\psi^{3})=\exp{[\pm\im \tilde{H}_{2}]}.		
	\label{ferm}
\end{aligned}
\end{equation}

The dimension of the left-moving vertex $V^L_{\vec{\mu}}$
\begin{equation}
\Delta^L(\vec{\mu})= \Delta_{gh}(q)+ \Delta_{ST} + \Delta_{CY}^L, 
\end{equation} 
where 
\begin{equation}
\begin{aligned}
&\Delta_{gh}(q)=-\frac{q(q+2)}{2},
\\
&\Delta_{ST}^L= \fr12 \vec{\lambda}^2,
\\
& \Delta_{CY}^L=\fr12 \vec{S}^2 + 
	\vec{p} \cdot \vec{q}+\fr12 \vec{p} \cdot \vec{a}^- + 
	\fr12 \vec{q} \cdot \vec{a}^+.
\end{aligned}
\end{equation}		
		
The  phase  $2\pi i \ \vec{\mu} \cdot \vec{\mu}$ of the mutual locality two  left-moving vertices
$V^L_{\vec{\mu}}(u)$ and $V^L_{\vec{\mu'}} (z)$ is equal 
\begin{equation}
\Delta^L(\vec{\mu}+\vec{\mu'})-\Delta^L(\vec{\mu})-\Delta^L(\vec{\mu'}). 
\end{equation}
It follows that
\begin{equation}
\vec{\mu} \cdot \vec{\mu}'= -qq' + \lambda \cdot \lambda' + \vec{S} \cdot \vec{S'}+ 
  \vec{p} \cdot\vec{q'} + \vec{p'} \cdot \vec{q}.
\end{equation}

The vector $\lm^{a}$ in the exponent $\exp( \im\lm^{a}\tilde{H}_{a})$ 
must satisfy the requirement of consistency with the structure $N=1$ on the left side, and for this  in the NS sector the vectors $\vec{\lm}$  fall into the classes $[0]$ and $[V]$, and in the R sector the vectors $\vec{\lm}$ fall into  the classes $[S]$ and $[C]$ of the weight lattice of $SO(1,3)$ algebra.

The integrality of $\vec{\mu} \cdot \vec{\mu}'$ is the condition of mutual locality of two vertices.
\section{Massless  left movers and N=1 Space-Time supersymmetry}\label{sec:6}

The vertices of massless physical states are interesting because they play two roles.The first is that it is the states of this set that must correspond to the observed elementary particles. 

Secondly, some of these vertices can be used to extend the symmetry of the theory. They can be taken as currents whose integrals become additional generators for this extension.

By imposing the requirements on the vertices to be simultaneously $Q_{BRST}$ and  $ D_{\vec{1},\vec{1}}$ cohomology, we find the following massless left vertices in the NS-subsector  with canonical picture number $(-1)$ and in the R-subsector sector with canonical picture number $(-\fr12)$.

In the NS subsector we obtain a left-hand vertex belonging to the $N=1$ Super-Poincaré vector representation of the space-time symmetry
\begin{equation}
	\exp(-\phi(z))\psi^{\nu}(z)\exp\left(\im p_{\mu}x^{\mu}(z)\right), 
	\quad
	\label{3.Vect}
\end{equation}
which can be rewritten using bosonization as
\begin{equation}
\exp\left(-\phi(z) + \im\lm^{a}\tilde{H}_{a}+\im p_{\mu}x^{\mu}(z)\right),
\end{equation}
where $\vec{\lambda}=(\pm 1,0)$ or $\vec{\lambda}=(0, \pm 1)$.

In the NS-subsector we also find   a left-hand vertex belonging to the $N=1$ Super-Poincaré singlet representation of the space-time symmetry
\begin{equation}
	\begin{aligned}
		&V_{\vec{m} }=\exp{(-\phi + \vec{m} \cdot \vec{X}^-+\im p_{\mu}x^{\mu}(z))},
		\\
		&V_{\vec{n}}=\exp{(-\phi(z) + \vec{n} \cdot \vec{X}^++\im p_{\mu}x^{\mu}(z))},
	\end{aligned}
\end{equation}
where $\vec{m} \in \Delta^+$ and $\vec{n} \in \Delta^-$.

In the Ramond subsector  with canonical picture number $(-\fr12)$ we find the vertices of the massless spinors
\begin{equation}
	\begin{aligned}
		&J^{\pm}(\vec{\sigma},\vec{S})=
		\exp\left(-\fr12\phi + 
		\im {\vec{\sigma}}\cdot \vec{\tilde{H}} +
		\im {\vec{S}\cdot \vec{H}} \pm \fr12
		(\vec{X}^+ \cdot \vec{a}^- - \vec{X}^- \cdot \vec{a}^+ )\right)
		\exp(\im p_{\mu}x^{\mu}(z)),
		\\ 
		&J^{\pm}(\dot{\vec{\sigma}},\dot{\vec{S}})=
		\exp\left(-\fr12\phi + 
		\im {\dot{\vec{\sigma}}}\cdot \vec{\tilde{H}} +
		\im {\dot{\vec{S}}\cdot \vec{H}} \pm \fr12
		(\vec{X}^+ \cdot \vec{a}^- - \vec{X}^- \cdot \vec{a}^+ )\right)
		\exp(\im p_{\mu}x^{\mu}(z)),
		\\ 
		&J^{\pm}(\dot{\vec{\sigma}},\vec{S})=
		\exp\left(-\fr12\phi + 
		\im {\dot{\vec{\sigma}}}\cdot \vec{\tilde{H}} +
		\im {\vec{S}\cdot \vec{H}} \pm \fr12
		(\vec{X}^+ \cdot \vec{a}^- - \vec{X}^- \cdot \vec{a}^+ )\right)
		\exp(\im p_{\mu}x^{\mu}(z)),
		\\ 
		&J^{\pm}(\vec{\sigma},\dot{\vec{S}})=
		\exp\left(-\fr12\phi + 
		\im {\vec{\sigma}}\cdot \vec{\tilde{H}} +
		\im {\dot{\vec{S}}\cdot \vec{H}} \pm \fr12
		(\vec{X}^+ \cdot \vec{a}^- - \vec{X}^- \cdot \vec{a}^+)\right)
		\exp(\im p_{\mu}x^{\mu}(z)),		
	\end{aligned}
\end{equation}		
where
\begin{equation}
    \begin{aligned}	
		&\sigma^{a}=\pm\fr12,  \   \sum_{a=1}^{2}\sigma^{a}=\pm 1, \ \ \ \
		\dot{\sigma}^{a}=\pm\fr12,  \ \sum_{a=1}^{2}\dot{\sigma}^{a}=0,		
	\end{aligned}
\end{equation}
and
\begin{equation}
\begin{aligned}	
		&S_i=\pm\fr12,  \   \sum_{i=1}^{5} S_i= \fr12,  \mod \  2  \ \ \ \
       \dot{S_i}=\pm\fr12, \  \ \sum_{i=1}^{5}\dot{S_i}=-\fr12,  \mod \  2  \ \ .	
\end{aligned}
\end{equation}

All of these vertices are BRST cohomology, but the ones that are also cohomology of $D_{\vec{m}}$ and $D_{\vec{n}}$ are only the following vertices
\begin{equation}
	\begin{aligned}
		&J^{+}_{\sigma}=
		\exp\left(-\fr12\phi + 
		\im {\vec{\sigma}}\cdot \vec{\tilde{H}} + \fr12 H^L_{CY}\right),
		\\ 
		&J^{-}_{\dot{\sigma}}=
		\exp\left(-\fr12\phi + 
		\im {\dot{\vec{\sigma}}}\cdot \vec{\tilde{H}} - \fr12 H^L_{CY}\right),
		\\ 
		&J^{+}_{\dot{\sigma}}=
		\exp\left(-\fr12\phi + 
		\im {\dot{\vec{\sigma}}}\cdot \vec{\tilde{H}} + \fr12 H^L_{CY}\right),
		\\ 
		&J^{-}_{\sigma}=
		\exp\left(-\fr12\phi + 
		\im {\vec{\sigma}}\cdot \vec{\tilde{H}}  - \fr12 H^L_{CY}\right),		
	\end{aligned}
\end{equation}	

where $H^L_{CY}(z)=\sum_{i=1} (i H_i - a^+_i X^{-}_i+ a^-_i X^{+}_i)$ and we have omitted the factors $\exp(\im p_{\mu}x^{\mu}(z))$.
\\

The first two currents $J^{+}_{\sigma}$ and  $J^{-}_{\dot{\sigma}}$ 
are mutually local, as are the other two currents. We will use  the first pair to extend Poincaré symmetry to $N=1$ space-time supersymmetry.

Further, the currents $J^{+}(\vec{\sigma},\vec{S})$ and 
$J^{-}(\dot{\vec{\sigma}},\dot{\vec{S}})$ for brevity will be called simply 
$J^{+}_{\sigma}$ and  $J^{-}_{\dot{\sigma}}$.

We select this  pair of currents to determine $N=1$ super-Poincaré supercharges as follows
\begin{equation}
	\begin{aligned}
		&{\cal{Q}}_{\sigma}= \oint  du  \  J^{+}_{\sigma}(u) =
		   \oint  du   \exp\left(-\fr12\phi + 
		\im {\vec{\sigma}}\cdot \vec{\tilde{H}} + \fr12 H^L_{CY}\right),
		\\ 
		&{\cal{Q}}_{\dot{\sigma}}= 
		\oint du \  J^{-}_{\dot{\sigma}} (u) =
		  \oint du  
		\exp\left(-\fr12\phi + 
		\im \dot{{\vec{\sigma}}}\cdot \vec{\tilde{H}}  - \fr12 H^L_{CY}\right). 
	\end{aligned}
	\label{3.LSUSYa}
\end{equation}

The supercharges $\cal{Q}_{\sigma}$ and ${\cal{Q}}_{\dot{\sigma}}$, which are spinors with respect to the Super-Poincaré algebra, form  together with the generators of this algebra, $P_\mu$ and $J_{\mu \nu}$,  the $N=1$ Poincaré superalgebra .
\\

In order to obtain $N=1$ space-time supersymmetry in the theory, we must leave from the vertices of $ V^{\vec{L}}_{\vec{\mu}_{L}} $,  where $\vec{\mu}_{L}=(q, \ \vec{\lambda}, \ Q^{L}_{CY})$, which are the cohomologies of $Q_{BRST}$ and  $ D_{\vec{1},\vec{1}}$, 
only those vertices that are mutually local with $J^{+}_{\sigma}$  
and $J^{-}_{\dot{\sigma}}$. 

These vertices are mutually local with the currents  $J^{+}_{\sigma}$ and $J^{-}_{\dot{\sigma}}$ if
\begin{equation}
	q + \sum_{a}\lm^{a}+ Q^{L}_{CY}	\in  2 \ \mathbb{Z}.
\end{equation}
where  $ Q^{L}_{CY}=\sum_i{S_i}+\vec{p} \cdot \vec{a}^- - \vec{q} \cdot \vec{a}^+ $ 
is the $U(1)$ charge of the Calabi-Yau subsector.
\\
This equation  is nothing more than the GSO condition for the vertices  in the left sector.

From the GSO equation  it follows that the total internal charges  $Q^{L}_{CY}$  of the vertices are integers or half-integers.
This can be obtained more explicitly if we consider that the vectors $\vec{\lm}$ are weights of the algebra $SO(1,3)$, belonging to one of the four conjugacy classes of the weight lattice. 

For the general case of the algebra $SO(2n)$, the weight lattice consists of four subspaces
\begin{equation}
	\begin{aligned}
		& (0): (0,0,0,...,0) + \text{any root};
		\\
		&  (V): (1,0,0,...,0)+ \text{any root};
		\\
		& (S):  (\frac{1}{2},\frac{1}{2},\frac{1}{2},...,\frac{1}{2})+
		\text{any root};
		\\
		& (C):  (-\frac{1}{2},\frac{1}{2},\frac{1}{2},...,\frac{1}{2})+
		\text{any root}.
		\label{3.class}
	\end{aligned}
\end{equation}
From   GSO equation it  follows that in NS sector the vectors  $\vec{\lm}$   fall into classes $[0]$ and $[V]$ and in R sector the vector $\vec{\lm}$ fall into classes$[S]$  and $[C]$ of $SO(1,3)$.
As a result, we obtain the following connection between the conjugacy class  
of  $\vec{\lm}$ and  the $U( 1)$ charge of the CY factor for all 4 cases of left vertices.

In the NS sector we get
\begin{equation}
	\begin{aligned}
		& Q^{L}_{CY} \in 2\mathbb{Z}+1, \ \   
		\vec{\lm}\in[0],
		\\
		& Q^{L}_{CY}\in 2\mathbb{Z}, \ \  \ \ 
		\vec{\lm}\in[V].
			\end{aligned}
	\label{3.LGSOsol}
\end{equation}

In the R sector we get				
		\begin{equation}
	\begin{aligned}
		& Q^{L}_{CY} \in 2\mathbb{Z}+\fr12, \ \ 
		\vec{\lm}\in[S],
		\\ 
		& Q^{L}_{CY} \in 2\mathbb{Z}-\fr12, \ \ 
		\vec{\lm}\in[C].
	\end{aligned}
	\label{3.LGSOsol-R}
\end{equation}

\section{Right-moving sector, N=0 CFT}\label{sec:7}

The space-time subsector of the right-moving sector with the central charge $4$ contains boson fields $\bar{X}^{\mu}(\bar{z})$.

In order to obtain a critical string with the total central charge of $26$ in the right-handed sector, we add the boson fields $Y_{I}(\bar{z})$, $I=1,...,8$, compactified onto the torus of the algebra $E(8)$ with a central charge of $8$ and the bosons $\Phi_{\al}(\bar{z})$, compactified onto the torus of the algebra $SO(10)$ with a central charge of $5$.
The final contribution to the critical dimension is given by the right-moving part of the compact Calabi-Yau subsector with  the central charge $9$.

The full right-moving energy-momentum tensor is 
\begin{multline}
	\bar{T}_{mat}(\bar{z})=\fr12(\eta_{\mu\nu}\bar{\d}\bar{X}^{\mu}\bar{\d}\bar{X}^{\nu}+
	(\bar{\d}Y_{I})^{2}+(\bar{\d} \Phi_{\al})^{2}+
	\times\\\times
	+\sum_{i=1}^{5}(\d \bar{X}^{+}_i \d \bar{X}^{-}_i+
	\frac{1}{2}(\d\Psi^{+}_i\Psi^{-}_i+\d\Psi^{-}_i\Psi^{+}_i)-
	\frac{1}{2} (a^+_i \d^2 \bar{X}^{-}_i + a^-_i \d^2 \bar{X}^{+}_i)).
\end{multline}

The general right-moving vertex can be written as
\begin{equation}\label{4.RVert}
	\begin{aligned}
		&\bar{V}^{\vec{l}}_{\vec{\mu}_{R}}(\bar{z})=
		P_{gh}(\bar{b},\bar{c})P_{st}(\bar{\d} \bar{X}^{\mu})
		P_{int}(\bar{\d}\bar{Y}^{I},\bar{\d}\Phi_{\al},\bar{H}_{i}, \bar{X}^+_{i}, \bar{X}^{-}_{i})
		\\
		&\exp{(\im \eta_{I}\bar{Y}^{I}+
			\im\Lm_{\al} \Phi^{\al}
		+	i\sum_{i=1}^5 S_i \bar{H}_i+\vec{p} \cdot \vec{\bar{X}}^-}+
		\vec{q} \cdot \vec{\bar{X}}^{+} + \im p_{\mu}x^{\mu}(\bar{z})),
	\end{aligned}
\end{equation}
where $\vec{\eta}$ is a vector of the $E(8)$ root  lattice, 
and $\vec{\Lm}$ is the vector of the $SO(10)$ weight lattice.

Similar to what was done in the left sector, we impose the requirements that the right moving vertices be both $BRST$ and 
$ D_{\vec{1},\vec{1}}$ cohomologies, and also respect symmetries under $E(8)$ and $E(6)$ symmetries.

As in the left-handed sector, the right-handed vertices of massless physical states are interesting because they play two roles. 
First, it is the states of this set that must correspond to observable elementary particles. 
Second, some of these vertices can be used to extend the symmetry of the theory. 
As a result, it turns out that, as in the left-hand sector, we find four types of massless vertices that satisfy our requirements. 
\\

 The first type $[0]$ consists of the following $SO(10)$ singlets (and of their $E(6)$ partners).
\\

It   includes  $SO(1,3)$ vector $\im\bar{\d}\bar{X}^{\mu}(\bar{z})$
and  the currents of $SO(10)$

\begin{equation}
	\begin{aligned}
		&J_{\al}(\bar{z})=\im\bar{\d}\Phi_{\al}(\bar{z}), \ \al=1,...,5,
		\\
		&J_{\vec{\rho}}(\bar{z})=
		\exp{[\im \rho_{\al} \Phi_{\al}]}(\bar{z}),
		\ \rho_{\al}=\pm 1, \ \sum (\rho_{\al})^{2}=2,
		\label{4.SO10}
	\end{aligned}
\end{equation}
where the vectors $\vec{\rho}$ are the roots of $SO(10)$
\begin{equation}
	\vec{\rho}=(\pm 1, \pm 1,0,0,0)+ \text{permutations};
\end{equation}

 It also includes the currents of $E(8)$ algebra
\begin{equation}
	J^{I}(\bar{z})=\im\bar{\d}\bar{Y}^{I}(\bar{z}), \ I=1,...,8,
	\
	J_{\vec{\eps}}(\bar{z})=\exp{[\im \eps_{I}\bar{Y}^{I}]}(\bar{z}), 
	\ (\vec{\eps})^{2}=2,
	\label{4.E8}
\end{equation}

\begin{equation}
	\begin{aligned}
		\vec{\eps}=
		\begin{cases}(\pm 1,\pm 1,0,0,0,0,0,0)+ \text{permutations},
			\\
			(\pm\frac{1}{2},...,\pm\frac{1}{2})+ \text{permutations},\, 
			\text{even  number  of}  +\frac{1}{2},
		\end{cases}
	\end{aligned}
\end{equation}
where the vectors $\vec{\eps}$ are the roots of $E(8)$ algebra.

The current of  $U(1)$ algebra  of the right Calabi-Yau sector
\begin{equation}
J_{CY}(\bar{z})=\bar{\d} \bar{H}_{CY}(\bar{z}), \ \ 
\bar{H}_{CY}(\bar{z})=\sum_{i=1} 
(i\bar{H}_i - a^+_i \bar{X}^{-}_i+ a^-_i \bar{X}^{+}_i)
\end{equation}
also belong to $[0]$.
\\

Also, there is another set of $[0]$: the massless right vertices, which also consist of BRST and Borisov cohomology and are $E(8)\times E(6)$ singlets depending on the Calabi-Yau data.

\begin{equation}\label{4.RVert}
	\begin{aligned}
& V (\vec{S},\vec{m},\vec{n}) = \ 
\exp{(i\sum_{i=1}^5 S_i\bar{H}_i+\vec{m} \cdot \vec{X}^-}+\vec{n} \cdot \vec{X}^{+} 
		+ \im p_{\mu}\bar{X}^{\mu})(\bar{z}),\\
	\end{aligned}
\end{equation}
whose dimension  
\begin{equation}
\Delta(\vec{S},\vec{m},\vec{n})=
\frac{\vec{S}^2}{2} + \vec{m} \cdot \vec{n}+
\frac{ \vec{m} \cdot \vec{a}^- +  \vec{n} \cdot \vec{a}^-}{2} =1,
\end{equation}
and $U(1)$  charges 
\begin{equation}
 Q_{CY}(\vec{S},\vec{m},\vec{n}) = 
\sum_iS_i - \vec{n} \cdot \vec{a}^+ + \vec{m} \cdot \vec{a}^- = 0. 
\end{equation}
Here $\vec{m}\cdot\vec{a}^-=\vec{n} \cdot \vec{a}^+ =1$, 
that is $\vec{m}\in \Delta^+$ and $\vec{n}\in \Delta^-$.

There are three cases of solutions to all the above requirements that are imposed on the $E(8)\times E(6)$ singlets.

The first case  is
\begin{equation}
V (\vec{S},\vec{m},\vec{n}) = \ 
\exp{(\vec{m} \cdot \vec{X}^- + \vec{n}\cdot \vec{X}^+
+ \im p_{\mu}\bar{X}^{\mu})(\bar{z})};
\end{equation}
where the vectors  $\vec{m}\in \Delta^+$, $\vec{n}\in \Delta^-$  and their product $\vec{m} \cdot \vec{n}$ is equal to $0$.

The second case of the  singlet of $SO(10)$ vertices is
\begin{equation}
V (\vec{S},\vec{m},\vec{n}) = \ \sum_i m_i
\exp{(-i \bar{H}_i+\vec{m} \cdot \vec{X}^{-}
+ \im p_{\mu}\bar{X}^{\mu})(\bar{z}))};
\end{equation}
where $\vec{m}\in \Delta^+$.
 
The third case is
\begin{equation}
V (\vec{S},\vec{m},\vec{n}) = \ \sum_i n_i
\exp{(i\bar{H}_i+\vec{n} \cdot \vec{X}^{+}
+ \im p_{\mu}\bar{X}^{\mu})(\bar{z}))}.
\end{equation}
where  $\vec{n}\in \Delta^-$.

The total number of singlet massless states depends on and is determined by the CY data.
In the case of  Quintic their number coincides with the result of D. Gepner \cite{Gep}.
\\

 The second type $[V]$ of the massless vertices consists of the following  vertices of $10$-dimensional vector  representation of $SO(10)$  (and their $E(6)$ partner).

\begin{equation} 
\begin{aligned}
&V^{[10]}_{\vec{m} }=\exp{( i \vec{\Lambda}\cdot \vec{\Phi} + \vec{m} \cdot \vec{X}^-)}.
\\ 
\end{aligned}
\end{equation}

 The third type $[S]$ consists of the following  vertices 
massless $SO(10)$ spinors (and their $E(6)$ partners)
\begin{equation}\label{right-moving-currents}
	\begin{aligned}
		& J^{+}_{\om}(\bar{z})=
		\exp{(\im \om_{\al}\Phi_{\al} + \fr12 \bar{H}_{CY})(\bar{z})}
		\\
		&\om_{\al}=\pm\fr12, \ \sum \om_{\al}=\fr12 \mod \ 2\mathbb{Z}.
	\end{aligned}
	\end{equation}

		The fourth  type $[C]$ consists of the following  vertices 
massless $SO(10)$ spinors (and their $E(6)$ partners).
		\begin{equation}\label{right-moving-currents}
	\begin{aligned}
		&J^{-}_{\dot{\om}}(\bar{z})=
		\exp{(\im \dot{\om}_{\al} \Phi_{\al} - \fr12 \bar{H}_{CY})(\bar{z})}, 
		\\
		&\dot{\om}_{\al}=\pm\fr12, \ \sum \dot{\om}_{\al}=-\fr12 \mod \ 2\mathbb{Z}.
	\end{aligned}
	\end{equation}

We also emphasize that  these currents are cohomologies of  the differential $ D_{\vec{1},\vec{1}}$. 
\\

Taking the currents $J^{+}_{\om}(\bar{z})$ and $J^{-}_{\dot{\om}}$ and adding them to the currents of the algebra $SO(10) \times U(1)$, we   extend the algebra $SO(10) \times U(1)$ to the algebra $E(6)$ and as  a result obtain explicit expressions for all $78$ of the algebra $E(6)$,
which is known to be considered  as a possible Grand unified gauge group.
 
It should be noted that the emergence of the $E(6)$ symmetry in the right sector, firstly, ensures the self-consistency of the theory in connection with the presence of $N=1$ supersymmetry in the left sector.
Secondly, it is both of these symmetries that are necessary for the Theory of Grand Unification.
\\

Note that  in the right sector, in addition to the vertex
$V^{[10]}_{\vec{m} }=\exp{( i \vec{\Lambda}\cdot \vec{\Phi} + 
\vec{n} \cdot \vec{X}^-)}$ which  belong to the type $[V]$
we also find massless vertices 
$V^{[1]}_{\vec{m}}$
and
$V^{[16]}_{\vec{m}}$
of the sixteen-dimensional $SO(10)$ representations.

These vertices are the $E(6)$ partners of $V^{[10]}_{\vec{m}}$,  
which also belong to the  representation $27$, where
\begin{equation} 
\begin{aligned}
&V^{[1]}_{\vec{m} }=\exp{(  -\bar{H}_{CY}+ \vec{m} \cdot \vec{X}^-)},
\\
\end{aligned}
\end{equation}

and  one more massless  vertex that belongs to the $16$-dimensional spinor representation of $SO(10)$
\begin{equation} 
\begin{aligned}
&V^{[16]}_{\vec{m} }=\exp{( i \vec{\omega} \cdot \vec{\Phi} - 
 \fr12 \bar{H}_{CY} +\vec{m} \cdot \vec{X}^-)},
\\ 
\end{aligned}
\end{equation}
where $\vec{m} \in \Delta^+$,
$\vec{\Lambda}=(\pm 1,0,0,0,0)+ \text{permutations}$. 

These three sets  of vertices are the cohomologies of the differentials $ D_{\vec{1},\vec{1}}$ form a 27-dimensional representation of E(6). 

The right massless sector also contains $\bar{27}$ representations of E(6), which are similarly constructed from dual representations of $SO(10)$ and whose vertices depend on $\vec{n} \in \Delta^-$ instead of  
$\vec{m} \in \Delta^+$ . 
The numbers of representations $27$ and $\bar{27}$ are equal to the number of points in the reflexive Batyrev polytopes for a given CY-manifold.

\section{Massless right movers and  E(6) gauge  symmetry }
\label{sec:8}
 
To obtain the $ E(8)\times E(6)$  gauge symmetry in the constructed model, we need to restrict the space of states in the right moving sector
to a set of states compatible with the action of the $E(8) \times E(6)$ generators.

This means that we have to choose and keep among the vertices that are BRST and $ D_{\vec{1},\vec{1}}$ cohomology, 
 only those that are mutually local  with the currents $E(8)$ and $E(6)$.
 
These requirements are met if the following "GSO" consistency equations are satisfied
\begin{equation}
	\begin{aligned}
	& \vec{\rho} \cdot \vec{\Lambda} \in  \mathbb{Z}, \ \ 
	\vec{\eps} \cdot \vec{\eta} \in \mathbb {Z} ,
	\\
	&\om\cdot\Lm + \fr12 Q^R_{CY}  \in \mathbb{Z}.
	\label{5.GSO}
	\end {aligned}
\end{equation}

From the "GSO" equations in the right-moving sector we  get that the $SO(10)$ parts of the solutions of the right vertices, that fall into one of the four conjugacy classes of the   $SO(10)$ weight lattice, determine the sixth 
Calabi-Yau component $Q^R_{CY}$ as follows
\begin{equation}
	\begin{aligned}
		&\vec{\Lm}\in[0]
		\Rightarrow Q^R_{CY}\in 2\mathbb{Z}, 
		\\
		&\vec{\Lm}\in[V]
		\Rightarrow Q^R_{CY} \in 2\mathbb{Z}+1,
		\\
		&\vec{\Lm}\in[S]
		\Rightarrow Q^R_{CY} \in 2\mathbb{Z}-\fr12,
		\\		
		&\vec{\Lm}\in[C]
	 \Rightarrow Q^R_{CY} \in 2\mathbb{Z}+\fr12.
	\end{aligned}
	\label{5.RGSOsol}
\end{equation}
The data of the above-considered vertices satisfy these conditions.

\section{Complete massless vertices in  explicit form}
\label{sec:9}

For phenomenological applications, the most important are the massless states of the Heterotic string, compactified to four dimensions.

In this section we explicitly represent complete vertices for massless physical states as products of suitable left and right vertices.
We will omit the factor $\exp\left(\im p_{\mu}x^{\mu}\right)$ in order to shorten notations.

In choosing suitable products of left and right vertices, we  follow D. Gepner's approach in \cite{Gep}
to achieve modular invariance in constructing heterotic string theory.

Namely, we  mutiply the left movers of [0] by the right movers 
of [V] and vice versa, and also mutiply the left movers of [S] by the right movers of [C] and vice versa.

As a result, we obtain only subsets of complete vertices belonging to each of the 4 types of massless physical states.
Other complete vertices, which are nothing more than partners of subset vertices, can be obtained by applying $N=1$ Super-Poincaré actions to the left and $E(8)\times(6)$ symmetries to the right vertices.

\subsection{The gravity  supermultiplets}

The first class of the massless physical states includes  the gravitational supermultiplet. 

These vertices are the vector representation of Space-Time 
$N=1$ Super-Poincare and the singlet representation of 
 gauge group $E(8)\times E(6)$.
 
\begin{equation}
\exp(-\phi(z)) \ \psi^{\mu}(z) \times \im\bar{\d}x^{\nu}(\bar{z}).
\quad
\label{3.Vect}
\end{equation}

\subsection{The  supermultiplets for the gauge  group $E(8)\times E(6)$ }

The second class includes   the adjoint representations of the  gauge groups 
$E(8)$ and  $E(6)$.

The  vertices  of $E(8)$ are
\begin{equation}
\begin{aligned}
	&V^{I}_{\mu}(z,\bar{z})= \exp(-\phi(z)) \ \psi^{\mu}(z) \times  
	\im\bar{\d}\bar{Y}^{I}(\bar{z}), \ I=1,...,8,
	\\
	& V_{\mu}^{\vec{\eps}}(z,\bar{z})=\exp(-\phi(z)) \ \psi^{\mu}(z) \times 
	\exp{[\im \eps_{I}\bar{Y}^{I}]}(\bar{z}),
	\label{4.E8}
	\end{aligned}
\end{equation}
where the vectors $\vec{\eps}$ are the roots of $E(8)$ algebra.

The vertices  belonging to adjoint representation of $SO(10)$  algebra
\begin{equation}
	\begin{aligned}
		&V_{\mu,\al}(z,\bar{z})=\exp\left(-\phi(z)\right) \, \psi^{\mu}(z) \times 
		\im\bar{\d}\Phi_{\al}(\bar{z}), \ \al=1,...,5,
		\\
		&V_{\mu,\vec{\rho}}(z,\bar{z})=\exp\left(-\phi(z)\right) \, \psi^{\mu}(z) \times 
		\exp{[\im \rho_{\al} \Phi_{\al}]}(\bar{z}),
		\label{4.SO10}
	\end{aligned}
\end{equation}
where the vectors $\vec{\rho}$ are the roots of $SO(10)$;

The vertex of  $U(1)$ algebra of the Calabi-Yau sector
\begin{equation}
\begin{aligned}
&V_{\mu}^{CY}(z,\bar{z})=\exp\left(-\phi(z)\right) \, \psi^{\mu}(z) \times 
\bar{\d} \bar{H}_{CY}(\bar{z}), 
\\ 
&\bar{H}_{CY}(\bar{z})=\sum_{i=1}  
(i\bar{H}_i - a^+_i \bar{X}^{-}_i+ a^-_i \bar{X}^{+}_i).
 \end{aligned}
\end{equation}

We then  extend the $E_8 \times SO(10)\times U(1)$ symmetry to $E_8 \times E_6$ 
using  $32$  spinor currents of  $SO(10)$ algebra  $J^{+}_{\om}$ and  $J^{-}_{\dot{\om}}$ given by \eqref{right-moving-currents}
\begin{equation}
\begin{aligned}
& J^{+}_{\om}(\bar{z})=
\exp{(\im \om_{\al}\Phi_{\al} + \fr12 \bar{H}_{CY})(\bar{z})}
\\
&\om_{\al}=\pm\fr12, \ \sum \om_{\al}=\fr12 \mod \ 2\mathbb{Z},
\\
&J^{-}_{\dot{\om}}(\bar{z})=
\exp{(\im \dot{\om}_{\al} \Phi_{\al} - \fr12 \bar{H}_{CY})(\bar{z})}, 
\\
&\dot{\om}_{\al}=\pm\fr12, \ \sum \dot{\om}_{\al}=-\fr12 \mod \ 2\mathbb{Z}.
\end{aligned}
\end{equation}
As a  result we obtain the $32$ additional  vertices  of $E(6)$ algebra
\begin{equation}
	\begin{aligned}
		&V_{\om}(z,\bar{z})=\exp(-\phi(z) \ \psi^{\mu}(z) \times J^{+}_{\om}(\bar{z}),
		\\
		&V_{\dot{\om}}(z,\bar{z})=\exp(-\phi(z) \ \psi^{\mu}(z) \times J^{-}_{\dot{\om}}(\bar{z}),	
		\label{4.SO10}
	\end{aligned}
\end{equation}
that together with the other $46$  vertices  of $SO(10)$ and $U(1)$ form the 
$78$-dimensional  adjoint representation of E(6) algebra.

The second  class also includes the superpartners of these vertices listed above.

\subsection{The supermultiplets $27$  and $\overline{27}$ of group $E(6)$}

The  third  class is the   representation $27$ of  algebra $E(6)$ . 
The set of its vertices includes a one-dimensional representation $SO(10)$
\begin{equation} 
\begin{aligned}
&  V_{\vec{m}}(z,\bar{z})=
\exp\left(-\phi + \vec{m} \cdot \vec{X}^-\right) 
\times 
\exp{( - \bar{H^R}_{CY}+ \vec{m} \cdot \vec{X}^-)},
\\
\end{aligned}
\end{equation}
the $10$-dimensional representation of $SO(10)$
\begin{equation} 
\begin{aligned}
& V_{\vec{\Lambda},\vec{m}}(z,\bar{z})=
\exp\left(-\phi + \vec{m} \cdot \vec{X}^-\right) 
\times 
\exp{( i \vec{\Lambda}\cdot \vec{\Phi} + \vec{m} \cdot \vec{X}^-)},
\\ 
\end{aligned}
\end{equation}
and  the 16-dimensional spinor representation of $SO(10)$
\begin{equation} 
V_{\vec{\om},\vec{m}}(z,\bar{z})=
\exp\left(-\phi + \vec{m} \cdot \vec{X}^-\right) 
\times 
\exp{\left( i \vec{ \om } \cdot \vec{\Phi} - 
 \fr12 \bar{H^R}_{CY} +\vec{m} \cdot \vec{X}^-\right)},
\end{equation}
where $\vec{m} \in \Delta^+$, $\vec{\Lambda}=(\pm 1,0,0,0,0)+ \text{permutations}$.

The condition $\vec{m} \in \Delta^+$ means that the number  of $27$ supermultiplets  
is determined by the number of  dots $\vec{m}$ of the Batyrev polyhedron corresponding to the Calabi-Yau manifold of the Heterotic string model under consideration.
Explicit expressions for the other vertices of the $27$ supermultiplet can be obtained by acting through OPE on these vertices by the generators of the $N=1$ Poincaré superalgebra.
Note that this operation does not break the mutual locality between the extended set of vertex operators that was between the original ones.

By similar actions we obtain the set of vertices of the 
$\overline{27}$ supermultiplet whose number is determined by the number  of dots $\vec{n}$ of the dual Batyrev polyhedron.

\subsection{The singlet supermultiplets of $E(8)\times E(6)$}

The fourth class   the massless states  is the vector representation of N=1 Super-Poincare algebra and   the singlet representation  of $E(8)\times E(6)$ algebra.

Its first set of these vertices is
\begin{equation}	
		V_{\vec{\lambda},\vec{m},\vec{n}}(z,\bar{z})=
		\exp\left(-\phi + 
		\im {\vec{\lambda}}\cdot \vec{\tilde{H}} 
		\right)
\times 
\exp(\vec{m} \cdot \vec{X}^-+\vec{n}\cdot \vec{X}^{+}),
\end{equation}
where the vectors $\vec{\lambda}=(\pm 1,0), (0, \pm 1) $,  $\vec{m}\in \Delta^+$, $\vec{n}\in \Delta^-$ 
and their product $\vec{m} \cdot \vec{n}=0$.

The second set is
\begin{equation}	
V_{\vec{\lambda},\vec{m}}(z,\bar{z})=
		\exp\left(-\phi + 
		\im {\vec{\lambda}}\cdot \vec{\tilde{H}} 
		\right)
\times  \sum_i m_i
\exp{(-i\bar{H}_i+\vec{m} \cdot \vec{X}^{-})},
\end{equation}
where the vectors $\vec{\lambda}=(\pm 1,0), (0, \pm 1) $ and $\vec{m}\in \Delta^+$.

The third set is
\begin{equation}
V_{\vec{\lambda},\vec{n}}(z,\bar{z})=
 	\exp\left(-\phi + 
		\im {\vec{\lambda}}\cdot \vec{\tilde{H}}
		\right)
\times \sum_i n_i
\exp{(i\bar{H}_i+\vec{n} \cdot \vec{X}^{+})},
\end{equation}
where $\vec{\lambda}=(\pm 1,0), (0, \pm 1) $ and  
$\vec{n}\in \Delta^-$.
	 
For the case when Calabi-Yau sector is defined by the Quintic polynomial we find a total number  of  $E(8)\times E(6)$ singlets, which is coincide  with the result in \cite{Gep} (see also \cite{BLT}).

\section{Conclusion}
In this paper we developed a method for explicitly constructing the models of Heterotic string compactified on the product of a torus of the Lie algebra  $E(8)\times SO(10)$ and  general Calabi-Yau manifolds of Berglund-Hubsch type.

The construction uses Batyrev-Borisov combinatorial approach together with the use of Vertex algebra of free bosonic and fermionic operators to construct the Vertex algebra of the CY sector.

We used the requirement for the simultaneous fulfillment of mutual locality of the left-moving vertices with the space-time symmetry generators and of right-moving vertices with generators of $E(8)\times E(6)$ gauge symmetry together with the requirement of mutual locality of complete (left-right) vertices of physical states. 

In particular, it is shown that the vertex operators for particles of the charged representation $27$ of the algebra $E(6)$ correspond to the elements of the reflexive Batyrev polyhedron $\Delta^+$, and for particles of the representation $\overline{27}$ to the elements of the dual polyhedron $\Delta^-$.

Also, it was also shown that the vertex operators for particles of the singlet representation of the $E(6)$ algebra include pairs of elements $\vec{m}$ and $\vec{n}$ of $\Delta^+$ and $\Delta^-$ polytopes whose pairing is zero, which allows one to calculate the number of singlets for a given Calabi-Yau manifold.

In particular, the total number of singlets in the Quintic case 
is 326, which coincides with the number obtained in \cite {Gep}.

In choosing suitable products of left and right vertices, we  follow D. Gepner approach \cite{Gep} to achieve the modular invariance in constructing heterotic string theory.
Although the correctness of this approach has been proved for the CY compactification  used in \cite{Gep}, it would be interesting to prove
the fulfillment of modular symmetry in the wider class of CY considered in this paper.

\section*{Acknowledgments} 
The author acknowledges   A. Litvinov, V. Batyrev, L. Borisov, D. Gepner, F. Malikov,  
S. Aleshin, I. Belavina, A. Elashvili, B. Feigin, G. Giorgadze,  M. Jibladze and D. Kazhdan  for  the helpful and interesting discussions. 
He is  grateful to  Weizmann Institute of Science and Joseph Meyerhoff Visiting Professorship for the hospitality  in the time while some of this work was done.  This work is supported by the Russian Science Foundation grant  23-12-00333. Also, the author would like to thank Grigory Makarov for the fruitful discussions, which resulted in the fixing of some inaccuracies in the text.

\end{document}